\documentclass[a4paper,12pt]{article}
\usepackage{amsfonts}
\usepackage{latexsym}
\usepackage{amssymb}
\date{}

\begin{document}

\title{Entropy Production and Equilibrium Conditions in\\ General-Covariant
Continuum Physics
}
\author{W. Muschik\footnote{Corresponding author:
muschik@physik.tu-berlin.de}\quad and\quad 
H.-H. von Borzeszkowski\footnote{ 
borzeszk@mailbox.tu-berlin.de}
\\
Institut f\"ur Theoretische Physik\\
Technische Universit\"at Berlin\\
Hardenbergstr. 36\\D-10623 BERLIN,  Germany}
\maketitle

            \newcommand{\be}{\begin{equation}}
            \newcommand{\ee}{\end{equation}\normalsize}
            \newcommand{\bee}[1]{\begin{equation}\label{#1}}
            \newcommand{\bey}{\begin{eqnarray}}
            \newcommand{\byy}[1]{\begin{eqnarray}\label{#1}}
            \newcommand{\eey}{\end{eqnarray}\normalsize}
            \newcommand{\R}[1]{(\ref{#1})}
            \newcommand{\C}[1]{\cite{#1}}
            \newcommand{\td}{{^{\bullet}}}
            \newcommand{\dm}{\diamond\!}
            \newcommand{\st}[2]{\stackrel{_#1}{#2}}


\abstract\noindent
Starting out with an entropy identity, the entropy flux, the entropy production
and the corresponding Gibbs and Gibbs-Duhem equations of general-covariant
conti\-nuum thermodynamics are established. Non-dissipative materials and 
equilibria are investigated. It is proved that equilibrium conditions only put
on material properties cannot generate equilibria, rather additionally, the
Killing property of the 4-temperature is a necessary condition for space-times
in which equilibria are possible.

\section{Introduction}

The special-relativistic version of Continuum Thermodynamics (CT) was founded
by Eckart \C{EC} in form of the special-relativistic theory of irreversible
processes. CT is based (i) on the conservation laws for the particle number
and the energy-momentum tensor and (ii) on the dissipation inequality and the
Gibbs fundamental law. In order to incorporate CT in (or, at least, to
harmonize it with) General Relativity, as a first step, one has to formulate
it on a curved space-time, i.e., to go over to its general-covariant
formulation. This step brings problems along that one has to solve before
taking Einstein's gravitational field equations into consideration. This paper
is devoted to some of these problems. In particular, it concerns questions as
to the entropy production, the Gibbs equation and the definition of
thermodynamic equilibrium.
\vspace{.3cm}\newline
As to thermodynamic equilibrium, vanishing entropy production is a necessary
condition to be satisfied. As is well-known \C{ST,CHBOR}, this can
be reached by two different requirements: Either by assuming that the
considered matter is a perfect fluid (then one need not impose any conditions
on the properties of the underlying space-time structure) or by requiring
space-time structures that allow for a Killing and a conform Killing vector
field, respectively (then one need not restrict the structure of the material).
\vspace{.3cm}\newline
To define thermodynamic equilibrium, in both cases, the condition of vanishing
entropy production has to be supplemented by further equilibrium
conditions. It should be mentioned that for the second case it was shown
\C{CHBOR} that, if the temperature 4-vector is a Killing vector, shear and
expansion of the material vanish, and furthermore, that by implying
Einstein's gravitational field equations, one can deduce most of the other
supplementary conditions defining an equilibrium. In the present paper,
the Killing equation and Einstein's equations are not exploited.
\vspace{.3cm}\newline
In the following, first we start out with an identity for the entropy 4-vector
\C{MB1,MB2}. This identity is split into the entropy production, the entropy
flux and the Gibbs equation which is connected with the entropy density by a
Gibbs-Duhem equation. As usual, the Gibbs equation allows the definition of a
state space which is here one of local equilibrium because of the special
choice of the entropy identity. 
\vspace{.3cm}\newline
Different forms of the entropy production are considered for discussing
non-dissi\-pa\-tive materials and equilibria. Non-dissipative materials
characterized by material-inde\-pen\-dent vanishing of entropy production are
rediscovered as perfect fluids \C{ST}. For defining equilibrium, beyond
vanishing entropy production additionally ``supplementary equilibrium
conditions'' are required \C{MB2} and are introduced ad hoc. A general result
is that in equilibrium the Killing property of the 4-temperature is valid for
all materials.

\section{Energy-momentum Tensor}

The energy-momentum tensor $\Theta^{ab}$ is the governing field of
GCCT\footnote{GCCT:\ 
\underline{G}eneral-\underline{C}ovariant \underline{C}ontinuum
\underline{T}hermodynamics}.
Its (3+1)-split is
\byy{P2c}
&&\Theta^{kl}\ =\ \frac{1}{c^4}eu^ku^l + \frac{1}{c^2}u^kq^l +
\frac{1}{c^2}q^ku^l + t^{kl},\qquad t^{kl}\ =\ t^{lk},
\\ \label{L0a}
&&t^{kl}\ =\ -p h^{kl} + \pi^{kl},\qquad \pi^{kl}h_{kl}\ =\ 0,\quad
t^{kl}h_{kl}\ =\ t^k_k\ =:\ - 3p,
\eey
with its components
\bee{J1}
e\ =\ \Theta^{kl}u_ku_l,\qquad q^j\ =\ \Theta^{kl}u_lh_k^j,\qquad
t^{ij}\ =\ \Theta^{kl}h_k^ih_l^j.
\ee
Here, the projector perpendicular to the 4-velocities $u^k$
and $u_i$ is
\bee{P7}
h^i_k\ =\ \delta^i_k - \frac{1}{c^2}u^iu_k.
\ee
The meaning of the (3+1)-components is as follows: energy density $e$, energy
flux density $q^j$, stress tensor $t^{ij}$, pressure $p$ and viscosity tensor
$\pi^{kl}$.

\section{Entropy Identity}
\subsection{The general case}

A continuum theory need beside the particle number balance which according to
Eckart \C{EC}\footnote{We consider especially one-component systems.} reads
\bee{P3}
N^k{_{;k}}\ =\ 0, \quad
N^k\ =\ \frac{1}{c^2}nu^k,\quad n\ :=\ N^ku_k,\quad (nu^k)_{;k}\ =\ 0,
\ee
and the energy-momentum balance equation
\bee{aP3}
\Theta^{kl}{_{;k}}\ =\ 0,
\ee
an entropy balance
\bee{P3a}
S^k{_{;k}}\ =\ \sigma + \varphi,\qquad S^k\ =\ \frac{1}{c^2}su^k + s^k
\ee
(particle flux density $N^k$, entropy 4-vector $S^k$, entropy production
$\sigma$, entropy supply $\varphi$).
The Second Law of Thermodynamics is taken
into account by the demand that the entropy production has to be non-negative
at each event for arbitrary materials
\bee{P4}
\sigma\ \geq\ 0.
\vspace{.3cm}\ee
For proceeding, we need the following 
\vspace{.3cm}\newline
$\blacksquare$\ \
{\sf Proposition}\C{MB1,MB2}: There exists an {\em entropy identity}
\bee{P5}
S^k \ \equiv\ \Big(s^k - \lambda q^k - \Lambda_m \Xi^{km}\Big) 
+ \Big(\mu N^k + \lambda u_l \Theta^{kl} + 2\Lambda_p h^p_mu_nS^{knm}\Big).
\ee
This identity contains quantities which stem from the (3+1)-splits of 
the energy-momentum tensor \R{P2c}$_1$, of the entropy
\R{P3a}$_{2}$, of the particle number flux density \R{P3}$_2$
and from the (3+1)-split of the spin tensor
\bee{L15b}
S^{kab}\ =\
\Big(\frac{1}{c^2}s^{ab} + \frac{1}{c^4}u^{[a}\Xi^{b]}\Big)u^k
+s^{kab} + \frac{1}{c^2}u^{[a}\Xi^{kb]}.
\ee
The scalar $\mu$ is defined by
\bee{P9a}
\mu\ :=\ \frac{1}{n}(s - \lambda e - \Lambda_m \Xi^m).\hspace{3cm}
\blacksquare\hspace{-4cm}
\ee
The two fields $\lambda$ and $\Lambda_p$ in \R{P5} can be chosen arbitrarily:
\R{P5} is an identity, that means, it is valid 
for arbitrary $\lambda$ and $\Lambda_p$. According to \R{P5}
and \R{P9a}, the part of $\Lambda_m$ which is
parallel to $u_m$ does not contribute.
The physical meaning of these two fields is determined by introducing below the
general-relativistic Gibbs equation and the entropy flux which bring
physics into the identity \R{P5}.

\subsection{Entropy identity of local equilibrium}

According to \R{P3a}$_1$, we obtain the sum of entropy supply and production by
calculating the divergence of \R{P5}. First of all the questions arises,
how to distinguish between entropy production and entropy supply? In
non-relativistic theories and here in thermodynamics on curved
spaces\footnote{without taking Einstein's equations into account}, this answer
is easy to give: external forces and moments generates an entropy supply
\C{MB2}. According to \R{aP3}, here no external forces and moments are taken
into account. Consequently, we have to cancel the entropy supply in \R{P3a}$_1$
\bee{P22}
\varphi\ \equiv\ 0.
\vspace{.3cm}\ee
Because the energy-momentum tensor $\Theta^{kl}$ is spin-dependent and the
spin appears explicitly in \R{P5} by the terms which are multiplicated with
$\Lambda_m$, the spin occurs twice in the entropy 4-vector. This two-fold
appearance of the spin can be specialized by the simplifying setting
$\Lambda_m\doteq 0$ in the entropy identity \R{P5}. As we will see below,
this setting generates a state space which belongs to local equilibrium. More
general state spaces need a $\Lambda_m$ which is different from zero. The
entropy identity \R{P5} becomes in this special case
\bee{P9a1}
S^k\ \equiv\ s^k -\lambda q^k + \frac{1}{n}(s-\lambda e)N^k +
\lambda u_l \Theta^{kl}.
\vspace{.3cm}\ee
According to \R{P3a}$_1$ and \R{P22}, the entropy identity \R{P9a1} results in
an other identity by differentiation
\bee{P13a}
\sigma\ \ =\ \Big(s^k -\lambda q^k\Big)_{;k} +
\Big[\frac{1}{n}(s-\lambda e)N^k\Big]_{;k} +
\Big(\lambda u_l \Theta^{kl}\Big)_{;k}.
\ee
If the entropy flux density $s^k$ and the entropy density $s$ are specified
in \R{P13a}, this identity receives a physical meaning, if additionally
the arbitrary field $\lambda$ is suitably chosen.
\vspace{.3cm}\newline
Using \R{P3a}$_2$ and \R{aP3}$_1$, \R{P13a} results in
\bey\nonumber
\sigma &=& \Big(s^k -\lambda q^k\Big)_{;k} + \Big(\frac{s}{n}\Big)_{,k}N^k
-\lambda\Big(\frac{e}{n}\Big)_{,k}N^k - \frac{e}{n}\lambda_{,k}N^k +
\\ \label{P12a}
&+&\Big(\lambda u_m\Big)_{;k}\Theta^{km}.
\eey
Multiplying \R{P2c} by $u_l$
\bee{L0}
u_l \Theta^{kl}\ =\ q^k + \frac{e}{n}N^k,
\ee
and taking \R{L0a} into account, the fifth term of \R{P12a} becomes
\bee{P13}
\Big(\lambda u_m\Big)_{;k}\Theta^{km}\ =\
\lambda_{,k}\Big(q^k + \frac{e}{n}N^k\Big) + 
\lambda u_{m;k}\Big(\frac{1}{c^2}u^kq^m -ph^{km} + \pi^{km}\Big).
\vspace{.3cm}\ee
Further, we have by use of \R{P7} and \R{P3}$_4$
\bee{L1b}
\lambda u_{m;k}ph^{km}\ =\ \lambda u^k{_{;k}}p =
\lambda p n\Big(\frac{1}{n}\Big)^\td,
\vspace{.3cm}\ee
Finally, the identity \R{P12a} is
according to \R{P13}, \R{L1b} and \R{P3}$_1$ results in
\bey\nonumber
\sigma &=& \Big(s^k -\lambda q^k\Big)_{;k} + \Big(\frac{s}{n}\Big)_{,k}N^k
-\lambda\Big(\frac{e}{n}\Big)_{,k}N^k -
\lambda p n\Big(\frac{1}{n}\Big)^\td +
\\ \nonumber
&+&\lambda_{,k}q^k + 
\lambda u_{m;k}\Big(\frac{1}{c^2}u^kq^m + \pi^{km}\Big)\ =\
\\ \nonumber
&=&\Big(s^k -\lambda q^k\Big)_{;k} + \frac{n}{c^2}\Big(\frac{s}{n}\Big)^\td
-\frac{n}{c^2}\lambda\Big(\frac{e}{n}\Big)^\td
-\lambda p n\Big(\frac{1}{n}\Big)^\td +
\\ \label{P12}
&+&\lambda_{,k}q^k + 
\lambda u_{m;k}\Big(\frac{1}{c^2}u^kq^m + \pi^{km}\Big).
\eey

\section{Entropy Production and gr-Gibbs Equation\label{GE}}

As already mentioned, the identity \R{P12} has to be transfered into the
expression for the entropy production by specifying the entropy flux $s^k$ and
the entropy density $s$. Obviously, \R{P12} contains terms of different kinds:
a divergence, time derivatives and two other terms. This fact gives rise to
the following definitions of the entropy flux and the gr-Gibbs equation
which generate the entropy production
\byy{G1}
s^k &:=& \lambda q^k \quad\longrightarrow\quad\lambda\ =\frac{1}{T},
\\ \label{G2}
\Big(\frac{s}{n}\Big)^\td &:=& \frac{1}{T}\Big(\frac{e}{n}\Big)^\td +
c^2 \frac{p}{T}\Big(\frac{1}{n}\Big)^\td,
\\ \label{aG2}
\sigma &=&
\lambda_{,k}q^k + 
\lambda u_{m;k}\Big(\frac{1}{c^2}u^kq^m + \pi^{km}\Big).
\eey
The second term of \R{aG2} is
\bey\nonumber 
&&\quad\lambda u_{m;k}\Big(\frac{1}{c^2}u^kq^m + \pi^{km}\Big)\ =\
\lambda u_{m;k}
\Big(\frac{1}{c^2}u^kq^m +\frac{1}{c^2}q^ku^m+ \pi^{km}\Big)\ =\ 
\\ \nonumber
&&=\ \Big[(\lambda u_{m})_{;k}-\lambda_{,k}u_{m}\Big]
\Big(\frac{1}{c^2}u^kq^m +\frac{1}{c^2}q^ku^m+ \pi^{km}\Big)\ =
\\ \label{bG2}
&&=\ (\lambda u_{(m}){_{;k)}}
\Big(\frac{1}{c^2}u^kq^m +\frac{1}{c^2}q^ku^m+ \pi^{km} \Big)-
\lambda_{,k}q^k.
\eey
Consequently, the entropy production \R{aG2} takes another shape
\bee{cG2}
\sigma\ =\ 
(\lambda u_{(m}){_{;k)}}
\Big(\frac{1}{c^2}u^kq^m +\frac{1}{c^2}q^ku^m+ \pi^{km} \Big).
\vspace{.3cm}\ee
For the squel, wee need another expression containing the (3+1)-components
of $\Theta^{km}$ which do not appear in \R{cG2}
\bey\nonumber
(\lambda u_{m})_{;k}\Big(\frac{1}{c^4}eu^ku^m -ph^{km}\Big)\ =\
(\lambda_{,k}u_m + \lambda u_{m;k})
\Big(\frac{1}{c^4}eu^ku^m -ph^{km}\Big)\ =\
\\ \label{dG2}
=\ \lambda_{,k}\Big(\frac{1}{c^2}eu^k\Big) - \lambda u_{m;k}ph^{km}\ =\ 
\frac{e}{c^2}\lambda^\td - \lambda pu^k{_{;k}}.\hspace{.5cm}
\eey
Summing up \R{cG2} and \R{dG2} results in another expression of the entropy
production
\bee{eG2}
\sigma\ =\ 
(\lambda u_{(m}){_{;k)}}\Theta^{km}
-\frac{e}{c^2}\lambda^\td + \lambda pu^k{_{;k}}.
\vspace{.3cm}\ee
We obtain a fourth form of the entropy production, if we decompose
as usual \C{N} the velocity gradient into its kinematical
invariants: symmetric traceless shear $\sigma_{nm}$, expansion $\Theta$, 
anti-symmetric rotation $\omega_{nm}$ and acceleration $\st{\td}{u}_n$:
\byy{aP27c}
u_{l;k}\ =\ \sigma_{lk} + \omega_{lk} + \Theta h_{lk}
+\frac{1}{c^2}\st{\td}{u}_l u_k,\hspace{2.cm}
\\ \label{P28}
\sigma_{lk}=\sigma_{kl},\ \omega_{lk}=-\omega_{kl},\quad 
u^l\sigma_{lk}=\sigma_{lk}u^k=u^l\omega_{lk}=\omega_{lk}u^k=0,
\\ \label{P29}
\sigma^k_{k}=\omega^k_{k}=0,\quad \Theta:=u^k_{k}.\hspace{1.8cm}
\eey
Consequently, the second term of \R{aG2} becomes with \R{L0a}$_2$ and \R{aP27c}
\bee{L13}
\lambda u_{l;k}\Big(\frac{1}{c^2}u^kq^l + \pi^{kl}\Big) \ =\ 
\lambda\Big(\frac{1}{c^2}\st{\td}{u}_l q^l+ 
\sigma_{lk}\pi^{kl}\Big),
\ee
and the entropy production \R{aG2} results in
\bee{U0}
\sigma\ =\ \lambda_{,k} q^k + 
\lambda\Big(\frac{1}{c^2} \st{\td}{u}_l q^l +
\sigma_{lk}\pi^{kl}\Big).
\vspace{.3cm}\ee
The question now arises, if the Pfaffian \R{G2} is an integrable one ? To
answer this question, we have to look for the entropy density belonging to the
gr-Gibbs equation \R{G2}. We start out with an ansatz for $s$ and prove that
it satisfies the gr-Gibbs equation \R{G2}, if a Gibbs-Duhem equation of the
intensitiv variables is valid. Starting out with
\bee{G2a}
s\ :=\ \frac{e}{T} + c^2\frac{p}{T},
\ee
and generating its differential
\bee{G2b}
\st{\td}{s}\ =\ \frac{1}{T}\st{\td}{e} + (e+c^2 p)\Big(\frac{1}{T}\Big)^\td +
\frac{c^2}{T}\st{\td}{p},
\ee
we demand as in thermostatics that the Gibbs-Duhem equation
\bee{G2c}
(e+c^2 p)\Big(\frac{1}{T}\Big)^\td + \frac{c^2}{T}\st{\td}{p}\ =\ 0
\ee
is valid for the intensitive variables. We obtain from \R{G2b} and \R{G2c}
\bee{G2d}
\frac{\st{\td}{s}}{n}\ =\ \frac{1}{T}\frac{\st{\td}{e}}{n}\ =\
\frac{1}{T}\Big(\frac{e}{n}\Big)^\td - \frac{e}{T}\Big(\frac{1}{n}\Big)^\td\
=\ \Big(\frac{s}{n}\Big)^\td - s\Big(\frac{1}{n}\Big)^\td
\ee
This results in the gr-Gibbs equation \R{G2} by use of \R{G2a}
\bee{G2e}
\Big(\frac{s}{n}\Big)^\td\ =\ \frac{1}{T}\Big(\frac{e}{n}\Big)^\td +
\Big(s-\frac{e}{T}\Big)\Big(\frac{1}{n}\Big)^\td\ =\
\frac{1}{T}\Big(\frac{e}{n}\Big)^\td +
c^2\frac{p}{T}\Big(\frac{1}{n}\Big)^\td,
\ee
that means, the entropy density \R{G2a} and the gr-Gibbs equation \R{G2}
are compatible with each other, if the Gibbs-Duhem equation \R{G2c} is
introduced.
\vspace{.3cm}\newline
According to the gr-Gibbs equation \R{G2e}, the {\em state space} is
spanned by the energy per particle and the volume per particle
\bee{L6k1}
\boxplus\ =\ \Big(\frac{e}{n}, \frac{1}{n}\Big).
\ee
This state space belongs to an one-component system in local equilibrium 
\C{LLL}. That is the reason why the identity \R{P9a1} generated from \R{P5} by
seting $\Lambda_m\doteq 0$ was called the ``entropy identity of local
equilibrium''. The constitutive quantities
\bee{L7a}
{\bf M}\ =\ (s/n,p,T,q^k,\pi^{km},\Xi^k, \Xi^{km}) 
\ee
are functions of the state space variables
\bee{L7b}
{\bf M}\ =\ {\cal M}(\boxplus),
\ee
which are called the {\em constitutive equations}\footnote{How to use the
constitutive equations in connection with the gravitational field equations
see \C{MB3}.}.
\vspace{.3cm}\newline
Special cases of space-times and materials are considered in the sequel.

\section{Non-dissipative Materials\label{NDM}}

A {\em non-dissipative material} is characterized by vanishing entropy
production
even in the case of non-equilibrium\footnote{Vanishing entropy production
is necessary, but not sufficient for equilibrium.} for all space-times.
Consequently by definition, all processes in non-dissipative materials are
reversible, and therefore these materials are those of thermostatics.
\vspace{.3cm}\newline
According to \R{aG2} and \R{cG2}, the following material parameters are
identical to zero for non-dissipative materials
\bee{U0a}
q^k_{\sf ndiss}\ \equiv\ 0,\quad 
\pi^{kl}_{\sf ndiss}\ \equiv\ 0.
\ee
According to \R{P2c} and \R{L0a}, the energy-momentum tensor is that of a
perfect fluid
\bee{U0b}
\Theta^{kl}_{\sf ndiss}\ =\ \frac{1}{c^4}eu^ku^l - ph^{kl}. 
\ee
Consequently, we rediscover \C{ST} the following\vspace{.3cm}\newline
$\blacksquare$\
{\sf Proposition:} Non-dissipative materials are characterized by vanishing
entropy production for arbitrary space-times, resulting in the
material conditions \R{U0a}: non-dissipative materials are perfect
fluids.\hfill$\blacksquare$
\vspace{.3cm}\newline
The vanishing entropy production does not generate equilibria for
non-dissi\-pa\-tive materials according to \R{dG2} which represents
an ``equation
of motion'' of reversible processes. An other fact results from \R{dG2}:
vanishing entropy production is necessary for euilibrium but not sufficient
for it, that means, we need beyond the vanishing entropy production additional
equilibrium conditions which are considered in the next section.

\section{Equilibrium\label{EQ}}

We start out with the question:
How are equilibrium and non-dissipative materials
related to each other ? Concerning non-dissipative materials, we are looking
for material properties generating vanishing entropy production for arbitrary
space-times independently of possible equilibria. 
Concerning equilibria, we are asking for those conditions which have to be
satisfied by the space-time and by the actual material properties in
equilibrium. Hereby, equilibrium is defined by {\em equilibrium
conditions} which are divided into necessary and supplementary ones \C{MB2}
which are idependent of each other
The necessary ones are given by vanishing entropy production and vanishing
entropy flux density
\bee{P23}
\fbox{$ 
\sigma^{eq}\ \st{\bullet}{=}\ 0,\qquad s^k_{eq}\ \st{\bullet}{=}\ 0.
$}
\ee
Independently, the
supplementary equilibrium conditions are given by vanishing material time 
derivatives, except for that of the 4-velocity
\bee{P24}
\fbox{$
\boxplus^\bullet_{eq}\ \doteq\ 0,\qquad\boxplus\
\neq\ u^l.
$}
\vspace{.3cm}\ee
From \R{L1b}$_3$ and \R{P24} follows
that the divergence of the 4-velocity vanishes in equilibrium for all materials
\bee{P25c}
u^k{_{;k}}{^{eq}}\ =\ 0 .
\ee
The necessary equilibrium condition \R{P23}$_2$ which stems from the entropy
flux density becomes according to \R{G1}
\bee{P25c1}
q^k_{eq}\ =\ 0 .
\ee
The equilibrium condition \R{P25c1} must not taken for
\R{U0a}$_1$ which represent an invariable
material property\footnote{the heat conduction coefficients are zero}, whereas
\R{P25c1} is only an actual one,valid in equilibrium for all
materials. In non-equilibrium we have $q^k\neq 0$ which is not true for
non-dissipative materials..   
The equilibrium conditions \R{P23} to \R{P25c1} are valid for all equilibria
and all materials.
\vspace{.3cm}\newline
We now investigate equilibrium in more detail. From \R{aG2} follows with
\R{P23}$_1$ and \R{P25c1}
\bee{P25c2}
u^{eq}_{m;k}\pi_{eq}^{kl}\ =\ 0\quad\longrightarrow\quad
\fbox{$
u^{eq}_{(m;k)}\ =\ 0\ \vee\ \pi_{eq}^{kl}\ =\ 0
$}\ ,
\ee
including \R{U0}
\bee{P25c3}
\sigma_{lk}^{eq}\pi_{eq}^{kl}\ =\ 0\quad\longrightarrow\quad
\fbox{$
\sigma_{lk}^{eq}\ =\ 0\ \vee\ \pi_{eq}^{kl}\ =\ 0
$}\ ,
\ee
and \R{cG2} results in
\bee{P25c4}
\fbox{$
(\lambda u_{(m})_{;k)}^{eq}\ =\ 0\ \vee\ \pi_{eq}^{kl}\ =\ 0
$}\ .
\ee
According to \R{P23}$_1$, \R{P24} and \R{P25c}, we obtain a sufficient and
necessary equilibriun condition from \R{eG2} for arbitrary materials
\bee{P26}
0\ =\ (\lambda u_{(m})_{;k)}^{eq}\Theta^{km}_{eq}\ =\ 
\Big[\Big(\frac{u_m}{T}\Big)_{;k}^{eq}+
\Big(\frac{u_k}{T}\Big)_{;m}^{eq}\Big]\Theta^{km}_{eq}.
\ee
Because the energy-momentum tensor $\Theta^{km}$ does not contain the
temperature, we obtain the Killing condition for the 4-temperature
$\lambda u_m$ as a general equilibrium condition
\bee{P27}
\fbox{$
0\ =\ 
\Big[
\Big(\lambda u_m\Big)_{;k}+\Big(\lambda u_k\Big)_{;m}
\Big]^{eq}
$}\ .
\ee
An interesting result is: Material properties subjected to equilibrium
conditions cannot generate equilibria. Additionally, the Killing property
\R{P27} is necessary characterizing the space-times which allows equilibria.
\vspace{.3cm}\newline
Finally, a remark on an expression which can be found in literature \C{ST,N}
may be useful: Subtracting \R{dG2} from \R{eG2} results in
\bee{P28a} 
\sigma\ =\ (\lambda u_{(m})_{;k)}\Big[\Theta^{km}-
\Big(\frac{1}{c^4}u^ku^m - ph^{km}\Big)\Big].
\ee
Here, $\sigma=0$ is not sufficient for equilibrium, because $\lambda^\td$ and
$u^k{_{;k}}$ can be different from zero in \R{dG2} and \R{eG2}. If equilibrium
is presupposed\footnote{that means, $\lambda^\td$ and $u^k{_{;k}}$ are zero
in \R{dG2} and \R{eG2}}, the material behaves as a perfect fluid independently
of the space-time, or if not, the space-time has to obey \R{P27}\footnote{or
the 4-temperature has to be a conform-Killing vector \C{CHBOR}}. This is in
accordance with the statement: $\sigma=0$ is necessary, but not sufficient for
equilibrium.

\section{Discussion}

Usually, relativistic thermodynamics starts out with a symmetric and
di\-ver\-gence-free energy-momentum tensor and an ansatz for the entropy
4-vector \C{N} whose divergence allows to find out the entropy production. The
shortcoming of such a procedure is that the initial ansatz contains no hint at
determining also the entropy flux fitting to that chosen entropy production. 
That is here the reason for starting out with an always valid entropy identity
\C{MB1,MB2} which allows to be split into three parts --entropy flux and
production and Gibbs equation-- which fit together. Here, a restricted entropy
identity is taken up, because only systems in local equilibrium are
investigated. Different forms of the entropy production are obtained,
emphazising energy transport and viscosity or shear, or showing off the
temperature 4-vector. Gibbs equation for entropy per particle number can be
defined unambiguously. The integrability of this Gibbs equation is confirmed
by a Gibbs-Duhem equality.
\vspace{.3cm}\newline
Non-dissipative materials are defined by vanishing entropy production
independently of the actual space-time: perfect fluids are rediscovered as the
only non-dissipative material \C{ST}. Although the entropy production is
identical to zero for all perfect fluids, there exist non-equilibria for them:
the reversible processes. This results in the fact that vanishing entropy
production is only necessary, but not sufficient for equilibrium.
\vspace{.3cm}\newline
Consequently, supplementary equilibrium conditions beyond the vanishing entropy
production are required for equilibrium: that are vanishing entropy flux and
vanishing time-space derivatives of the thermodynamic quantities. This results
in the well known fact that the 4-temperature is a Killing vector in
equilibrium for all materials.
\vspace{.3cm}\newline
Although not all results of this paper are really new, the method of deriving
them is strict and does not depend on an ansatz of the entropy 4-vector. Beyond
that, the used procedure can be easily extended to spin systems out of local
equilibrium and to systems in General Relativity taking Einstein's field
equations into account.


\begin{thebibliography}{99}

\bibitem{EC}
C. Eckart: Thermodynamics of irreversible proceses III, Phys. Rev. 58
(1940) 919

\bibitem{ST}
H. Stephani: General Relativity. An introduction to the theory of the
gravitational field, Cambridge University Press, Cambridge 1982, 2nd edition
1990

\bibitem{CHBOR}
T. Chrobok, H.-H. v. Borzeszkowski: Thermodynamical equilibrium and spacetime
geometry, Gen Relativ Gravit 38 (2006) 397-415

\bibitem{MB1} W. Muschik, H.-H. v. Borzeszkowski: Entropy identity and
material-independent equilibrium conditions in relativistic thermodynamics,
arXiv:0804.2659v1 [gr-qc] 16 Apr 2008

\bibitem{MB2} W. Muschik, H.-H. v. Borzeszkowski: Entropy identity and
equilibrium conditions in relativistic thermodynamics. Gen Relativ Gravit
41 (2009 1285-1304

\bibitem{LLL} W. Muschik, C. Papenfuss, H. Ehrentraut: A Sketch of
Continuum Thermodynamics. J. Non-Newtonian Fluid Mech. 96 (2001)
255-290, Sect.3.4.

\bibitem{MB3} W. Muschik, H.-H. v. Borzeszkowski: Exploitation of the
Dissipation Inequality in General Relativistic Continuum Thermodynamics.
Arch. Appl. Mech. 84 (2014) 1517-1531 

\bibitem{N} G. Neugebauer: Relativistische Thermodynamik, Vieweg Braunschweig
(1980) ISBN 3-528-06863-9



\end{thebibliography}
\end{document}